\documentstyle[11pt,newpasp,twoside,epsf,rotate]{article}
\def\edcomment#1{\iffalse\marginpar{\raggedright\sl#1\/}\else\relax\fi}
\reversemarginpar

\begin{document}
\title{The Magnificent Seven: Close-by Cooling  Neutron 
Stars ?}
\author{Aldo Treves}
\affil{Dipartimento di Scienze, Universit\`a  dell'Insubria,
Via Valleggio 11, 22100, Como, Italy; e--mail: treves@mi.infn.it}
\author{Sergei B. Popov}
\affil{Sternberg Astronomical Institute, Universitetskii Pr. 13,
119899, Moscow, Russia; e--mail: polar@sai.msu.ru}
\author{Monica Colpi}
\affil{Dipartimento di Fisica, Universit\`a di Milano Bicocca,
P.zza della Scienza 3, 20126 Milano, Italy; e--mail: colpi@uni.mi.astro.it}
\author{Mikhail E. Prokhorov}
\affil{Sternberg Astronomical Institute, Universitetskii Pr. 13,
119899, Moscow, Russia; e--mail: mystery@sai.msu.ru}
\author{Roberto Turolla}
\affil{Dipartimento di Fisica, Universit\`a di Padova, Via
Marzolo 8, 35131 Padova, Italy; e--mail: turolla@pd.infn.it}

\begin{abstract}

We model Galactic populations of accreting and cooling isolated neutron 
stars in the attempt to explore their link with
a new class of dim soft X-ray sources revealed by ROSAT. 
For accretors we follow the magneto-rotational and dynamical evolution in the 
Galactic potential and a realistic
large scale distribution of the interstellar medium is used.
Under standard assumptions old neutron stars enter the accretor stage
only if their magnetic field exceeds $\approx 10^{11}$--$10^{12}$ G. 
We predict about 1 source per square degree for fluxes 
$\approx 10^{-15}$--$10^{-16}$
erg cm$^{-2}$s$^{-1}$ in the energy range 0.5-2 keV.

Cooling neutron stars are explored within a simpler model of local 
sources, including however interstellar absorption. They are found to be 
significantly less abundant at low fluxes, $<0.1$
sources per square degree, but dominate over accretors at higher flux levels
($\approx 10^{-12}$--$10^{-11}$ erg cm$^{-2}$s$^{-1}$).
We suggest that the faint sources observed by ROSAT may be  
young cooling neutron stars with typical age  $\la 10^6$ yrs, 
if the total
number of young neutron stars in the Solar proximity is $\sim 10$ 
times higher 
than inferred from radiopulsars statistics. 

\end{abstract}

\section{The Magnificent Seven}

Seven soft 
sources, {\it The Magnificent Seven}, 
have been found in ROSAT fields which are most 
probably associated with isolated radio-quiet NSs. A summary of their
main observational properties is reported in the table  (see e.g. 
Ne\"uhauser \& Tr\"umper \cite{nt99}, NT99; Motch \cite{mo2000}; Treves et 
al. \cite{t2000}, T2000, and references therein; see also Burwitz et al. 
at this conference).

\begin{table}
\label{table1}
\caption{Properties of ROSAT Isolated NS Candidates}
\bigskip
\begin{tabular}{lccccc} 
Source& PSPC & $T_{bb}$ & $N_H$ & $\log f_X/f_V$ & Period \\
& count$\, \rm{s}^{-1}$ & eV & $10^{20}\, {\rm cm}^{-2}$ &  & s\\

 MS 0317.7-6647 & 0.03 & 200 & 40 & $> 1.8$ & -- \\
 RX J0420.0-5022 & 0.11 & 57 & 1.7 & $>3.3$ & 22.7 \\
 RX J0720.4-3125 & 1.69 & 79 & 1.3 & 5.3 & 8.37 \\ 
 RX J0806.4-4132 & 0.38 & 78 & 2.5 & $> 3.4$ & -- \\ 
 RBS1223 & 0.29 & 118 & $\sim 1$ & $> 4.1$ &   5.2 \tablenotemark{a}
\tablenotetext{a}{Hambaryan et al. (2001)} \\
 RBS1556 & 0.88 & 100 & $< 1$ &$ > 3.5$ & -- \\
 RX J185635-3754 & 3.64 & 57 & 2 & 4.9 & --  \\ 

\end{tabular}
\end{table}
 
Present X-ray and optical data however do not allow an unambiguous
identification of the physical mechanism responsible for their emission. 
These sources
can be powered either by accretion of 
the interstellar gas onto old ($\approx
10^{10}$ yr) NSs or by the release of 
internal energy in relatively young ($\approx
10^6$ yr) cooling NSs (T2000 
for a recent review). The ROSAT
candidates, although relatively bright 
(up to $\approx 1 \ {\rm count\,s}^{-1}$), are
intrinsically dim and their inferred luminosity 
($L\approx 10^{31} \ {\rm erg\, s}^{-1}$)
is near to that expected from either a close-by cooling NS or 
from an accreting NS among the most luminous.
Their X-ray spectrum
is soft and thermal, again as predicted for 
both accretors and coolers (T2000).
Up to now only two optical counterparts have been identified: RXJ 1856-37,
Walter \& Matthews (\cite{wm97}), for which a distance estimate of $\sim
60$ 
pc has been very recently obtained,  and RXJ 0720-31, 
Kulkarni \& Van Kerkwick (\cite{kk98}). In both cases an
optical excess over the low-frequency tail of the black body X-ray spectrum 
has been reported.

A statistical approach, based on the comparison of the predicted and observed
source counts may provide useful informations on the nature of these
objects. Previous studies derived the $\log N$ -- $\log S$
distribution of accretors 
(Treves \& Colpi \cite{tc91}; Madau \& Blaes \cite{mb94};
Manning et al. \cite{mann96}) 
assuming  a NSs velocity distribution rich in slow
stars ($v\la 100 \ {\rm km\, s}^{-1}$). 
Recent measurements of pulsar velocities 
(e.g. Lyne \& Lorimer \cite{ll94}; Hansen \& Phinney \cite{hp97})
and upper limits on the observed number 
of accretors in ROSAT surveys (Danner
\cite{dan98}) point, however, to a larger NS mean velocity
(T2000). Ne\"uhauser \& Tr\"umper (NT99)
compared the number count distribution of the
ROSAT isolated NS  candidates  with those of accretors and coolers.
Here we address these issues in  greater detail, 
in the light of the latest contributions to the modeling of the evolution
of Galactic NSs (Popov et al. \cite{p2000a}, P2000a). A more 
comprehensive discussion can be found in Popov et al. (\cite{p2000b}, 
P2000b).  

\section{Accreting isolated neutron stars}

An important feature of our approach is the detailed calculation
of the magneto-rotational evolution of isolated NSs, calculated as in P2000a
(see Lipunov \cite{l92} for the basic concepts),
but with slightly revised values for the critical periods governing 
the Ejector-Propeller and Propeller-Accretor
transitions.
When the accretor stage is reached, the NS spin
period is set equal to the ``equilibrium'' period
(Konenkov \& Popov \cite{kp97}). 
The form of the Galactic potential is taken  as in Paczynski
(\cite{pac90}; see also P2000a); some 
parameters were upgraded in order to fit better
solar distance from the Galactic center
(see Madau and Blaes \cite{mb94}).
Initially each NS has a circular velocity corresponding to its
birthplace, and an additional kick velocity, selected from a Maxwellian
distribution, is added. NSs are assumed to be born in the Galactic plane and 
their birthrate is constant in time and proportional to the the square
of the 
local gas density.
The NS magnetic field is taken to follow a log-gaussian distribution. 
For each evolutionary track we calculate six different magneto-rotational 
histories, corresponding to different values of the initial magnetic field. 
Results were then merged and normalized. 
The total number of NSs in the Galaxy is taken $N_{tot}=10^9$. 

In order to compare theoretical predictions with observations it is useful 
to produce the $\log N$ -- $\log S$ distribution. The brightest accretors 
in our calculations
have luminosities $L\sim 10^{32}$ erg s$^{-1}$, but
the majority of them cluster around $L = 10^{29}-10^{30}$ erg s$^{-1}$.
To compare our results with observations of ROSAT sources a conversion factor
$0.01\, {\rm count} \, {\rm s} ^{-1} =3\times 10^{-13} {\rm erg} \, {\rm cm}^{-2}
{\rm s}^{-1}$ was used (e.g. NT99).
We calculate the $\log N$ -- $\log S$  distribution
for the frequency intergrated flux ($S_{total}=L_{total}/4\pi D^2$, 
here $D$ is the 
source distance, $L_{total}= \dot M GM/R$) and for the 
flux in the  range 0.5-2 keV.
In latter case the spectrum is assumed to be a blackbody and the polar cap
radius is calculated with the current values of the magnetic field and 
accretion rate ($R_{cap}=R_{NS}\sqrt{R_{NS}/R_A}$, where $R_A$ is the
Alfven radius).
If absorption is negligible, then one expects about 1 source per square
degree in the  range 0.5-2 keV for limiting fluxes about
$10^{-16}$--$10^{-15}$ erg cm$^{-2}$ s$^{-1}$. This should be compared 
with a total numer of sources presently detected by Chandra of 
$\sim 10^3$--$10^4$ (see e.g. Giacconi at this conference).
Results are presented in the figure.

\section{Cooling neutron stars}

Although in principle the evolution of coolers should be computed exactly in 
the same way as for accretors, they are much more short-lived and hence
less numerous. This poses a severe problem about the reliability of our 
sample which is based on a limited number of evolutionary tracks. This can 
be avoided exploiting the fact that coolers are a local population of sources.
We assume that NSs are uniformly distributed in the disk with  
half-thickness of 450 pc. The NSs spatial density is a free parameter and 
can be varied.
We used two values, $n_{NS}=0.33\times 10^{-3} {\rm pc}^{-3}$ and 
$n_{NS}=3.3\times 10^{-3} {\rm pc}^{-3}$. The first value corresponds
to the density adopted by NT99, and comes from radiopulsars statistics.
The second one corresponds to 
$N_{tot}\sim 10^9$ and is suggested by considerations on 
supernova nucleosynthetic
yields (e.g. Arnett et al. \cite{ar89}). 
These two values can be compared, for example, with $n_{NS}\sim 1.4\times 
10^{-3}$pc$^{-3}$ as derived by Paczynski (\cite{pac90}). 

All NSs are assumed to be ``standard candles'' with $L=10^{32}$ erg s$^{-1}$
and blackbody spectrum. The duration of the cooling phase was taken to be 
$10^6$ yrs, as suggested by the slow cooling scenario. 
In the fast cooling model (e.g. Yakovlev et al. \cite{yak99}) the number 
of observable
coolers should be much smaller, so, potentially, observations of isolated 
NSs may help in shedding light on their cooling history.
The ISM structure is treated in a very simple way: a spherical local Bubble 
of radius $r_l$ (which can be varied)
and density $n=0.07$ cm$^{-3}$ centered at the Sun and
a uniform medium  with density $n=1$ cm$^{-3}$ in the Galactic disc 
(with scaleheight 450 pc) at larger distances. 
After the column density $N_H$ is calculated, we 
compute the source count rate for given luminosity, 
temperature and column density. Results are shown in the figure.

\begin{figure}[h]
\epsfxsize=0.6\hsize
\centerline{\rotate[r]{\epsfbox{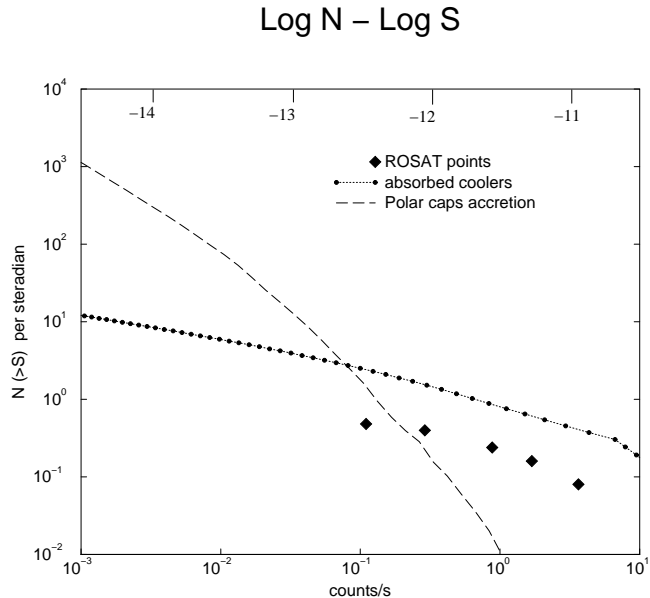}}} 
\caption{Comparison of the $\log N$ -- $\log S$ distributions for 
accretors and coolers.
Top and bottom axes give the flux in erg cm$^{-2}$s$^{-1}$ and in ROSAT 
counts/s.
\label{fig1}}
\end{figure}

This simple model reproduces a key feature: the flattening of the 
$\log N$ --$\log S$ distribution outside the Local Bubble, which is important
to explain ROSAT data at large fluxes.
More sophisticated models with a realistic distribution of both 
the ISM and the coolers give nearly the same results, and  
will be presented in a separate paper. 
In the actual calculation we took $r_l=140$ pc (equal to the radius of 
the Local Bubble in our calculations for accretors, see also Sfeir et al. 
\cite{sfeir99}). A clear knee appears due to the effect of absorption.
Such strong flattening can help to explain the observed data,
if one assumes that the the vast majority  of bright ($>0.1$ count s$^{-1}$)
isolated NSs are  already identified. 
The position of the knee can be adjusted varying $r_l$,
$n_{NS}$ and the ISM density;
for smaller values of $r_l$ the knee moves to the right, towards higher 
count rates.
By taking a high enough spatial density of NSs these results are able to 
account for both: i) the number of bright sources, and ii) 
the flattening of the observed  $\log N$ -- $\log S$ distribution.

\section{Discussion}

The task of explaining the observed properties of the seven ROSAT 
isolated NS candidates within a single model based on standard assumptions is 
not easy. 
The ROSAT sources are in 
fact: i) relatively bright, $\ga 0.1$ count s$^{-1}$; ii) close, 
$N_H\sim 10^{20} {\rm cm}^{-2}$;
iii) soft, $T_{eff}\sim 50 - 100$ eV; iv) slowly rotating, for
RX J0720, RX J0420 and RBS1223 the periods are about 5--20 s. 

The accretion stage  can be reached only if NS magnetic field is
$\ga 10^{11}$--$10^{12}$ G. For polar cap
accretion the X-ray spectrum is then relatively hard with 
a typical temperature around
300-400 eV. For them interstellar absorption is not very significant,
and we predict about 1 source per square degree for fluxes 
about $10^{-15}$--$10^{-16}$ erg cm$^{-2}$ s$^{-1}$ 
for energy range 0.5-2 keV.
While accretors may indeed represent the bulk of a still 
undetected low-luminosty population of X-ray sources, it is
difficult to reconcile them with the relatively bright ROSAT sources, as 
shown in the figure.   
On the contrary, the number distribution of coolers is in agreement 
with that of the seven isolated NS sources discovered so far, 
only if the total number of
young close-by NSs is a factor $\sim 10$ larger than what implied by 
radiopulsar 
statistics. Taken at face value, this implies that the majority of NSs do not
experience an active radiopulsar phase, a major implication which is
also supported by other evidences (see e.g. Gotthelf \& Vasisht 
\cite{gv2000}).  
The interpretation of the observed periods of some of the Magnificent Seven remains
problematic and seems to require special
conditions on the evolution of the magnetic field and spin.  
\bigskip
 
\noindent
{\bf Acknowledgements}
The work of SBP and MEP was supported through the grant RFBR 00-02-17164. 
Support from the European
Commission under contract ERBFMRXCT98-0195 and from MURST under contract
COFIN98021541 is acknowledged.

\end{document}